# An Analysis of Research in Software Engineering: Assessment and Trends

Zhi Wang [a,b], Bing Li [c,d], Yutao Ma [b,d,*]

a State Key Lab of Software Engineering, Wuhan University, Wuhan 430072, China

b School of Computer, Wuhan University, Wuhan 430072, China

c International School of Software, Wuhan University, Wuhan 430079, China

d Research Center for Complex Network, Wuhan University, Wuhan 430072, China

**Abstract:** Glass published the first report on the assessment of systems and software engineering scholars and institutions two decades ago. The ongoing, annual survey of publications in this field provides fund managers, young scholars, graduate students, etc. with useful information for different purposes. However, the studies have been questioned by some critics because of a few shortcomings of the evaluation method. It is actually very hard to reach a widely recognized consensus on such an assessment of scholars and institutions. This paper presents a module and automated method for assessment and trends analysis in software engineering compared with the prior studies. To achieve a more reasonable evaluation result, we take into consideration more high-quality publications, the rank of each publication analyzed, and the different roles of authors named on each paper in question. According to the 7638 papers published in 36 publications from 2008 to 2013, the statistics of research subjects roughly follow power laws, implying the interesting Matthew Effect. We then identify the Top 20 scholars, institutions and countries or regions in terms of a new evaluation rule based on the frequently-used one. The top-ranked scholar is Mark Harman of the University College London, UK, the top-ranked institution is the University of California, USA, and the top-ranked country is the USA. Besides, we also show two levels of trend changes based on the EI classification system and user-defined uncontrolled keywords, as well as noteworthy scholars and institutions in a specific research area. We believe that our results would provide a valuable insight for young scholars and graduate students to seek possible potential collaborators and grasp the popular research topics in software engineering.

**Keywords:** Systems and Software Engineering; Assessment; Trends Analysis; Research Publications; Power Law

[*] Corresponding author. Tel.: +86 27 68776081
E-mail: {zhi.wang (Z. Wang), bingli (B. Li), ytma(Y.T. Ma)}@whu.edu.cn

# 1 Introduction

Scientific research is a primary mechanism by which a discipline (or a field) attempts to accomplish its advances. In order to better understand where the discipline (or field) in question has been, and to consider where it may be going, the analysis of research conducted within it has been widely recognized as a reasonable and feasible method [1], mainly including assessment and trend analysis.

For a specific discipline (or research field), such a method presents its history and current status and predicts future directions through the statistics on a large number of papers published in peer-reviewed journals, which provides various audiences with important reference for different purposes. For example, an assessment of scholars, institutions and countries (or regions) is valuable to evaluate the performance of research institutions and their scholars in a quantitative and comprehensive way [2], while the trend analysis for a certain research field is of importance to those newcomers who are seeking for future research directions and possible collaborative research opportunities [3].

Software engineering is a relatively new research field derived from computer science. Over six decades, from 1948 until today, its importance has been widely recognized by more and more scholars within the field of computing, and it becomes an active and promising subdivision of the computing field. Like other disciplines, such as cancer [4], agriculture [5] and geographic information system [6], the assessment and trend analysis have long been applied to software engineering [7], but there are several problems that remain unsolved [2].

(1) Since there are only seven journals selected as the result of a survey at most, the size of samples (i.e., the number of referred papers published in these journals) is small, implying that the results may be one-sided.

(2) Because the keywords analyzed were collected from the Top 15 scholars to best describe their research focus, they are likely to be subjective and biased, which may not be used to reasonably reflect the trends and hot topics in software engineering.

(3) The scoring schemes for leading scholars and institutions were designed using the evaluation rule (see Section 2) proposed in [7], which overlooks the leadership role of few scholars among all authors of a multiple-authored paper.

To the best of our knowledge, the latest paper of the annual survey of publications in systems and software engineering from 1994 hasn't been published till now, despite few of reports on the subdivisions of software engineering such as agile software development [8]. Thus, the main goal of this paper is twofold: on one hand, we will present a new assessment of scholars, institutions and countries (or regions) in software engineering from 2008 to 2013, as well as a survey of trend analysis of this field over the past six years; on the other hand, a more reasonable and general method for assessment and trend analysis, which overcomes the above-mentioned existing problems in prior studies, will be proposed to accomplish such a study with more publications than ever before.

Furthermore, it is worth to note that the study in this paper is actually based on empirical evidence, that is, the results may rely mainly on the data analyzed. To reduce data errors and ensure the repeatability of our results, we selected 24 prestigious journals and 12 famous international conferences (research track) in systems and software engineering, and obtained author list, institution list, keyword list and other information of each paper under discussion from the Elsevier EI (Engineering Village) Compendex database[1].

In summary, the main contributions of this study are described as follows.

(1) In addition to an assessment of software engineering scholars and institutions (2008-2013), this paper also presented two interesting results, namely, noteworthy scholars and institutions in a specific research field as well as popular research trends, based on a large sample of 7638 research papers published in 36 different publications during this period.

(2) In consideration of the different roles of authors of a paper, this paper proposed a new evaluation rule for scholars, institutions and countries (or regions) based on the one [7] frequently used in prior studies; furthermore, our method for assessment and trend analysis was implemented by a software program, thus leading to automated data processing rather than manual operation.

(3) We found that the distributions of scholars, institutions, countries (or regions) and keywords roughly followed power laws in terms of their corresponding scores, and theoretically proved that small data errors (e.g., few of papers are missing) of our method have hardly any impact on the Top 15 (or even 20) ranking results.

The rest of this paper is organized as follows. Section 2 is a review of related literature. Section 3 presents the method for assessment and trend analysis to address the research questions. Section 4 introduces the data analyzed, experimental procedure and experimental results in detail. Section 5 discusses some important issues associated with this study, as well as some threats to validity that could affect our study. Finally, Section 6 concludes the paper and presents the agenda for future work.

## 2  Related Work

In 1994, Robert L. Glass released the first report on the assessment of systems and software engineering scholars and institutions [7]. Based on the number of research papers published in six equally-ranked leading journals, namely, IEEE Transactions on Software Engineering, ACM

---
[1] http://www.engineeringvillage.com

Transactions on Software Engineering and Methodologies, IEEE Software, Software: Practice and Experience (John Wiley & Sons), Journal of Systems and Software (Elsevier), and Information and Software Technology (Elsevier), his study attempted to answer the following two questions: (1) *Who are the most published scholars in the field of systems and software engineering*? (2) *Which are the most published institutions*?

In order to calculate the score of each scholar in question, he defined an evaluation rule based on counting schemes used in a prior similar study of evaluating authors in the field of information systems [9]. A single author of a published paper receives a score of one, while each author of a multiple-authored paper initially receives a score equal to their fractional representation on the paper; for author totals, the initial scores for multiple authors are updated with the following transformation: 0.5 becomes 0.7, 0.33 becomes 0.5, and those values that are less than or equal to 0.25 become 0.3. On the other hand, an author's raw scores (without the transformation) are attributed to the institution he or she belongs to on a paper. Thus, the Top 15 scholars and institutions in systems and software engineering field were ranked in terms of such scores.

Subsequently, such studies have been published each year. All of them [7, 10-12], released between 1994 and 1997, only examined the short-term publication data (i.e., 1, 2, 3, and 4 years were covered). In [13], Robert L. Glass began to carry out such a study with five years' worth of data. And then, all of the following studies [14-22, 2] covered the most recent five year period, so as to identify with more confidence those top scholars and institutions in this field.

In [16], keywords were considered by such studies for the first time. The authors sent an E-mail to each of the Top 15 scholars and asked them to provide a set of keywords which can best describe their research focus within the study period. For the keywords collected from those scholars, the analysis indicated their diversity and similarity of research topics to some extent. At the same time, the authors first investigated the geographic distribution of the Top 15 scholars, and they found that most of the Top 15 scholars were from the United States of America (five), the Asia-Pacific region (five) and Europe (four).

In 2009, a new publication—Empirical Software Engineering (Springer)—was added to the journal list of the thirteenth study in the series [22], because the authors wanted to emphasize the importance of applied software engineering research with a strong empirical component. Since then, the number of publications referred by the follow-up study [2] has increased to seven. As of June 2014, the latest result of this study hasn't been released when we finished our experiments. Nevertheless, similar assessments of scholars and institutions in the subdivisions of software engineering, such as empirical software engineering [23] and agile software development [8], have recently been released to provide useful information for various researchers who are interested in specific research topics in software engineering.

The series of such studies does have practical limitations, despite its success and profound impact. Since modern software engineering is inherently human-centric, Buse *et al.* tried to find a more effective assessment method based on user evaluation [24], and they identified trends about, benefits from, and barriers to performing such a method in software engineering research. The result indicated that citation counts were correlated with the presence of user evaluations.

Wohlin conducted a series of studies on article citations in software engineering [25-28], and he presented a list of the 20 most cited articles in this field according to the papers published in 20 famous journals. The author believed that his studies could provide valuable insights into what research to focus on now and in the future. Furthermore, Hummel *et al.* [29] analyzed citation indices (such as the *h*-index) of almost 700 researchers in the field of software engineering, and they illustrated citation values of world-class scholars, e.g., the top *h*-index scores in software engineering are around 60. This could provide a useful approach for sponsors of research activities to assess the work of scholars they (plan to) support.

Besides the assessment of scholars and institutions, mining software engineering popular research topics and trends is also an interesting problem. Cai *et al.* [30] examined 691 papers published in 7 top journals and 7 top international conferences in order to answer the following question: *what are the active research focuses within the field of software engineering*? Based on the ACM Computing Classification System (CCS), the result showed that Testing and Debugging (D.2.5), Software/Program Verification (D.2.4) and Management (D.2.9) were the Top 3 hot topics in 77% of journal or conference papers analyzed.

Unlike the trend analysis of software engineering based on a systematic literature review, Hoonlor *et al.* [31] investigated the entire research field of computer science in the past two decades by using the frequency of occurrence of keywords on those publications in the ACM Digital Library[2] and the IEEE Xplore Digital Library[3]. For example, for the data collected from the ACM Digital Library, they used the ACM CCS and author-defined keywords to respectively study the broader and static versus the finer and dynamic views of the trends in computer science. The study showed overall trends that provide a clear picture of the direction each topic is taking, and general rules of the evolution of researchers and research communities. However, because the ACM Digital Library and the IEEE Xplore Digital Library have different topic classification systems, this may affect the completeness and correctness of the results.

# 3 Research Questions and Method

## 3.1 Research Questions

In order to comprehensively present the research status and trends of software engineering

---
[2] http://dl.acm.org
[3] http://ieeexplore.ieee.org

from 2008 to 2013, this paper aims to investigate the following five research questions.

- *RQ1: Who are the most published scholars*? The goal of *RQ1* is to identify those scholars who can be worthy of attention in terms of the high-quality papers published for their leadership or participation during the period.

- *RQ2: Which are the most published institutions*? The goal of *RQ2* is to find those prestigious research institutions according to their staff members' outstanding studies in the field of software engineering.

- *RQ3: Which are the most published countries (or regions)*? The goal of *RQ3* is to distinguish those leading countries or regions on the basis of their competitive research institutions' total contributions in this field.

- *RQ4: What are the most popular research topics and trends*? The goal of *RQ4* is to mine those hot topics and popular trends in software engineering in with the light of formal classification terms and user-defined keywords on the papers in question.

- *RQ5: Which scholars and institutions are considered particularly noteworthy for a specific research topic in software engineering*? According to the results of *RQ1*, *RQ2* and *RQ4*, the goal of *RQ5* is to recommend the most published scholars and institutions together from the perspective of research focus.

## 3.2  Method

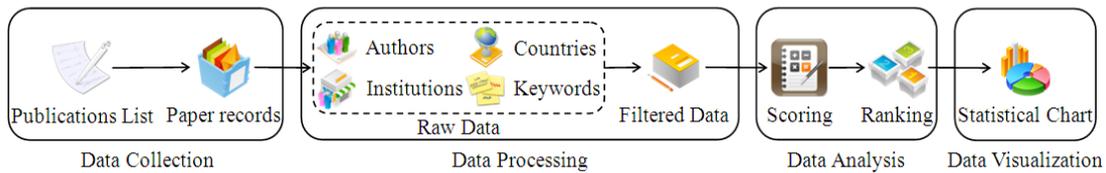

Figure 1. General framework of the method

As shown in Figure 1, our method has a standard process composed of four sequential modules, namely, data collection, data processing, data analysis and data visualization, which are automatically implemented by a software program.

First, according to those chosen publications, the software program collects all paper records within the specified period of time from the EI Compendex database, and the details please refer to the subsection 3.2.1.

Second, it garners the raw data about experimental subjects, namely, author, institution, country/region and keyword, from all paper records obtained. In consideration of duplication of names (especially Chinese names), different abbreviations of an author affiliation, and same meaning of multiple controlled and user-defined keywords, we filter out the raw data and store the filtered data as four key-value hash lists, where every instance in each hash list (e.g., *Robert L. Glass*, *Stanford University*, *USA* and *Computational Complexity*) is unique. Note that, in this paper we identify an institution from both academia and industry without respect to its branches and departments.

Third, the software program gives every instance of each experimental subject a score according to our evaluation rule (see the subsection 3.2.2), and it returns their corresponding Top 20 rankings to answer the questions *RQ1*, *RQ2*, *RQ3* and *RQ4*. For the *RQ5*, we re-calculate the scores of those leading scholars and institutions using a small part of the filtered data for a special research topic. Note that, the procedure that counts the score of each experimental subject is independent of each other. That is, for example, the score of an institution is not the total sum of its affiliated scholars' scores, because a scholar may move from one university to other universities during the period analyzed.

Finally, the software program presents the statistical charts of experimental results by calling data visualization tools such as Microsoft Excel and the R Project for Statistical Computing[4].

### 3.2.1 Data Collection

As mentioned before, our empirical study is data-driven, implying that data collection is an important phase and lays a foundation for the following assessment and trend analysis. Based on Figure 1, the detailed process of data collection is shown in Figure 2.

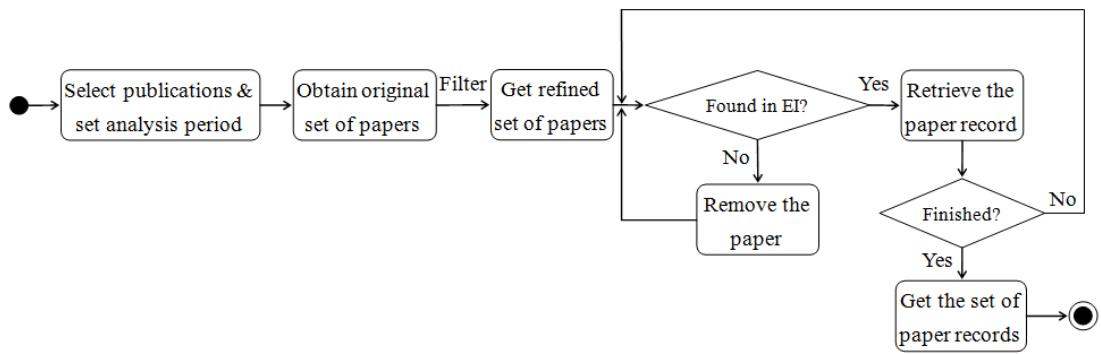

Figure 2. Process of data collection

Unlike the series of annual survey of publications in systems and software engineering [7, 10-22, 2], our method examines the research papers published in 36 different publications, including 24 journals and 12 conferences, shown in Table 1. The grades of these journals and conferences are ranked by the China Computer Federation (CCF)[5] according to their quality and prestige in the field of systems and software engineering.

---

[4] http://www.r-project.org
[5] http://www.ccf.org.cn/sites/ccf/biaodan.jsp?contentId=2567814757404

Table 1. Brief introduction to the publications analyzed

| Publication Name | Publication score |
|---|---|
| Journals:<br>● Class A<br>ACM Transactions on Programming Languages and Systems (TOPLAS), ACM Transactions on Software Engineering and Methodology (TOSEM), IEEE Transactions on Software Engineering (TSE)<br>● Class B<br>Automated Software Engineering (ASE, *Springer*), Empirical Software Engineering (ESE, *Springer*), IEEE Transactions on Services Computing (TSC), IET Software (IETS), Information and Software Technology (IST, *Elsevier*), Journal of Functional Programming (JFP, *Cambridge University Press*), Journal of Software: Evolution and Process (JSEP, *Wiley*), Journal of Systems and Software (JSS, *Elsevier*), Requirements Engineering (RE, *Springer*), Science of Computer Programming (SCP, *Elsevier*), Software and System Modeling (SoSyM, *Springer*), Software: Practice and Experience (SPE, *Wiley*), Software Testing, Verification and Reliability (STVR, *Wiley*)<br>● Class C<br>Computer Languages, Systems and Structures (CLSS, *Elsevier*), International Journal on Software Engineering and Knowledge Engineering (IJSEKE, *World Scientific*), International Journal on Software Tools for Technology Transfer (STTT, *Springer*), Journal of Logic and Algebraic Programming (JLAP, *Elsevier*), Journal of Web Engineering (JWE, *Rinton Press*), Service Oriented Computing and Applications (SOCA, *Springer*), Software Quality Journal (SQJ, *Springer*), Theory and Practice of Logic Programming (TPLP, *Cambridge University Press*) | Journals:<br>● Class A: 10<br>● Class B: 5<br>● Class C: 2 |
| Conferences:<br>● Class A<br>ACM SIGSOFT Symposium on the Foundation of Software Engineering / European Software Engineering Conference (FSE/ESEC, *ACM*), Conference on Object-Oriented Programming Systems, Languages, and Applications (OOPSLA, *ACM*), International Conference on Software Engineering (ICSE, *ACM/IEEE*), USENIX Symposium on Operating Systems Design and Implementations (OSDI, *USENIX*), ACM SIGPLAN Symposium on Programming Language Design & Implementation (PLDI), ACM SIGPLAN-SIGACT Symposium on Principles of Programming Languages (POPL), ACM Symposium on Operating Systems Principles (SOSP)<br>● Class B<br>European Conference on Object-Oriented Programming (ECOOP, *AITO*), IEEE International Requirement Engineering Conference (RE), International Conference on Automated Software Engineering (ASE, *ACM/IEEE*). International Conference on Software Maintenance (ICSM, *IEEE*), International Symposium on Software Testing and Analysis (ISSTA, *ACM*) | Conferences:<br>● Class A: 6<br>● Class B: 3 |

As we know, the DBLP (Digital Bibliography & Library Project) Computer Science Bibliography[6] provides free online access to its database and distinguishing marks for the table of contents of each publication included (especially conference proceedings), but the information of every paper it provides is too simple. On the contrary, because the EI Compendex database is not a free digital database, the access to its records through the terminal of a legal institution subscriber

---
[6] http://www.informatik.uni-trier.de/~ley/db

is always restricted by the institution, though the information of an EI paper record is very rich.

Therefore, an original set of papers published in the target publications within the analysis period is collected first from the DBLP by our software program through an open and easy-to-use interface. We then remove non-research papers such as editorial, industry track paper and short research paper from the original set, with the help of the annotations provided by the DBLP. For those papers in the refined set, the software program call the tool *Octopus Collector*[7] to retrieve their corresponding records, including title, author, author affiliation, corresponding author, publisher, publication year, classification code, controlled terms, etc., from the EI Compendex database using batch mode. Note that, any paper will be deleted from the refined set if the tool doesn't return its corresponding record in the EI Compendex database. Finally, we obtain the data set of papers to be analyzed in our experiments after the procedure of data collection ends.

### 3.2.2 Counting Schemes

Quantifying an author's contributions in terms of his/her referred papers published in the chosen publications, is the most significant component of our method. In order to provide a more reasonable assessment, our method adopts a new evaluation rule, namely comprehensive research evaluation rule (denoted by *C* rule), with an assumption of the differences among the publications in question.

Take the example of an assessment of authors. The *C* rule, based on the widely used rule [7] in prior studies, takes into consideration the authorship of all the papers that a scholar has published, which is suitable to be used as an evaluation criterion for the total research contributions of a scholar. However, the contributions of different authors of a paper are actually uneven, because the primary research of a paper is always dominated by few scholars. That is, the first author and the corresponding author play in general a leadership role in completing a paper.

To scholars:

- The single author of a paper receives a score of the publication that contains the paper (called *basic score*).
- For multiple-authored papers, both the first author and the corresponding author receive the same basic score if they are not the same person (otherwise he/she receives the basic score only once), while the other authors receive half of the basic score respectively.

In fact, a paper may be completed by multiple scholars from different research institutions, and the first author and the corresponding author are from the same institution in most cases. If there is more than one institution named on a paper, we must distinguish their sequential orders, namely, the first institution, the second institution, and so on. As we know, according to the rule for writing scientific papers, the name of a country/region is followed by the name of an institution located in the country/region. In some cases, if one country/region appears more than once in a paper, we only consider the first occurrence of such a name.

---

[7] http://www.bazhuayu.cc

To institutions (the rule for countries or regions is the same):

- The single institution named on a paper receives the basic score.
- For the papers completed by multiple institutions, the first institution receives the basic score, while the other institutions receive half of the basic score respectively.

Each EI Compendex record provides both standard classification terms and user-defined keywords. We argue that these EI classification terms are objective and valuable to analyze popular research topics and their changes, compared with those keywords collected from the Top 15 scholars in prior studies or from the ACM/IEEE Digital Library. However, the impacts that the same keyword occurs on the papers published in different publications on professional readers may be different. For example, if the term *model checking* appears on a paper published in the IEEE TSE, we tend to believe that it is more representative than those occurring in unknown journals or conferences.

To keywords:

- A keyword receives the basic score for the paper that contains the keyword.

## 4  Experimental Results and Findings

### 4.1  Statistics of Experimental Results

After performing the whole process of our method, we collected 7638 EI paper records, which contain more than 14 thousand authors, more than four thousand six hundred institutions, about 200 countries (or regions), and more than 6 thousand keywords. For the keywords, we further classified these keywords obtained into two types, namely, macro-keyword and micro-keyword. The macro-keywords represent the standard keywords defined by the EI Compendex database, e.g., classification terms, which reflect macro-level research subfields in software engineering. The micro-keywords denote those user-defined or uncontrolled keywords, which imply micro-level research topics.

In this paper, we utilized four frequently-used functions, namely, exponential function, polynomial function, logarithmic function and power function [32], to fit the curves of the scores of sorted scholars, institutions, countries/regions and keywords. As shown in Table 2, the distributions of the scores of experimental subjects except macro-keyword are best described by power laws, suggesting that only a few of leading scholars or institutions do receive much higher scores than those in the long tail. The finding implies that the Matthew effect also exists in software engineering research, and it highlights the assessment of top scholars and institutions as well as the trend analysis in this field.

Table 2. Fitting functions for different experimental subjects

| Subject | Logarithmic | Polynomial | Exponential | Power |
|---|---|---|---|---|
| Scholar | $y=-9.67\ln(x)+83.84$ ($R^2=0.917$) | $y=3\text{E-}06x^2-0.017x+33.76$ ($R^2=0.708$) | $y=22.90e^{-0.05x}$ ($R^2=0.856$) | $y=702.1x^{-0.58}$ ($R^2=0.968$) |
| Institution | $y=-12.6\ln(x)+100.1$ ($R^2=0.570$) | $y=4\text{E-}06x^2-0.024x+33.01$ ($R^2=0.336$) | $y=12.50e^{-0.03x}$ ($R^2=0.786$) | $y=5412x^{-1.03}$ ($R^2=0.964$) |
| Country/Region | $y=-620.\ln(x)+3052$ ($R^2=0.409$) | $y=0.065x^2-21.48x+1518$ ($R^2=0.249$) | $y=243.7e^{-0.02x}$ ($R^2=0.788$) | $y=11029x^{-1.97}$ ($R^2=0.961$) |
| Micro-keyword | $y=-34.7\ln(x)+255.1$ ($R^2=0.581$) | $y=6\text{E-}05x^2-0.153x+100.4$ ($R^2=0.357$) | $y=49.83e^{-0.01x}$ ($R^2=0.761$) | $y=1669x^{-0.68}$ ($R^2=0.992$) |
| Macro-keyword | $y=-247.\ln(x)+1222$ ($R^2=0.557$) | $y=0.028x^2-9.158x+647.7$ ($R^2=0.371$) | $y=358.5e^{-0.02x}$ ($R^2=0.951$) | $y=37262x^{-1.63}$ ($R^2=0.919$) |
| Note that, $Y$, $X$, and $R^2$ denote score, ranking, and degree of fitting, respectively. | | | | |

## 4.2 Leading Scholars

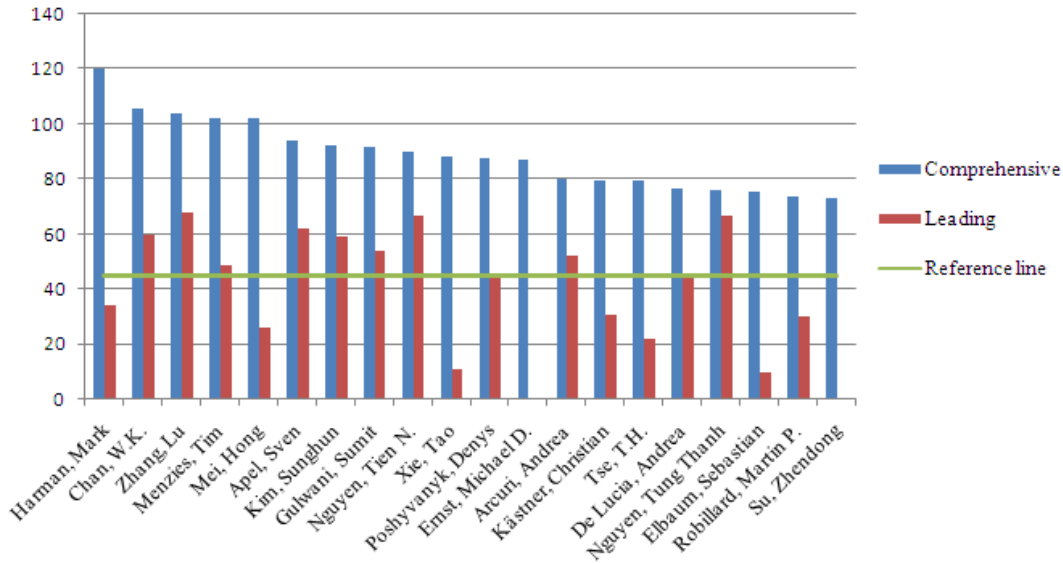

Figure 3. Top 20 scholars ranked in terms of the *C* rule

The scores of all scholars were calculated in terms of the *C* rule, and the Top 20 ranking list of scholars is shown in Figure 3, where Y-axis denotes the score that a scholar receives and the legend *Comprehensive* represents the scores calculated based on the *C* rule. As shown in Figure 3, these leading scholars have achieved a score of 73 or more during the years covered in this study. Professor Mark Harman of the University College London tops the list with a score of 120, Dr. Wing-Kwong Chan of the City University of Hong Kong is runner-up, and Professor Lu Zhang of the Peking University finishes third.

In order to make a comparison between the comprehensive contributions and the dominated contributions of leading scholars, we calculated the scores for their leadership according to a simple rule, namely, only the first author or the corresponding author of a multiple-authored paper receives the basic score, while the other authors receive a score of zero. In Figure 3, such scores are denoted by the legend *Leading*. Surprisingly, few well-known scholars in the list receive very low scores for their leadership, largely because these scholars omit marking their status of corresponding author in a great many papers they published.

Then, we also drew a reference line, which represents the first 20[th] scholar's score for his/her leadership, to visualize such a comparison for each scholar. Although nine out of 20 scholars received high scores in terms of the *C* rule, their scores for leadership are lower than the reference line with a score of 45, indicating that leading scholars in software engineering do not seem to attach much weight to the status of corresponding author of a paper, compared with those scholars in other disciplines such as medical science, biology and physics.

## 4.3 Leading Institutions

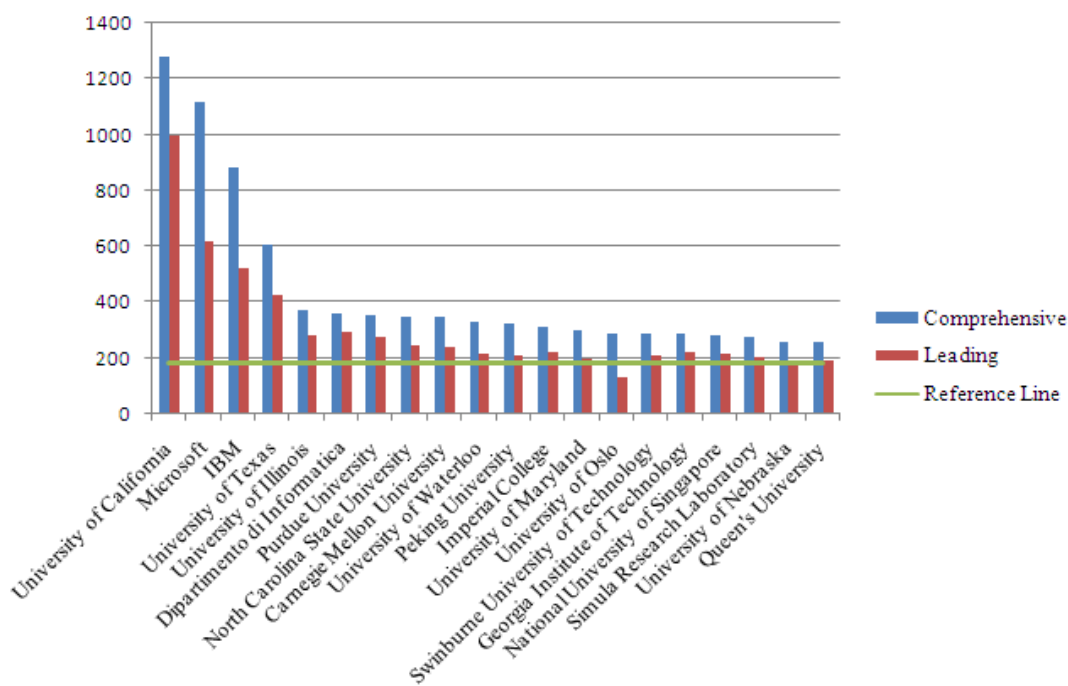

Figure 4. Top 20 institutions ranked in terms of the *C* rule

The Top 20 institutions list is shown in Figure 4, where Y-axis represents the score that an institution receives. The meanings of the legends *Comprehensive*, *Leading* and *Reference Line* are the same as those defined in Figure 3. It is worth noting that only the first institution named on a paper receive the basic score for its leadership. As shown in Figure 4, these leading institutions have achieved a score of 257 or more during the years covered in this study, and the scores of the first four institutions are much higher than those of the others of the Top 20, implying that it follows the "twenty-eighty Rule".

University of California (UC), Microsoft and IBM take first, second and third place, respectively. The champion is a public university system in the USA, which has 10 famous campuses like UC Berkeley; unexpectedly, the runner-up and the second runner-up are world-class IT corporations, perhaps because both of the corporations have several research labs around the world. In contrast with leading scholars, the vast majority of the Top 20 institutions (except the University of Oslo) exceeded or reached the reference line for their leadership, suggesting that these institutions did dominate the studies of the papers they published.

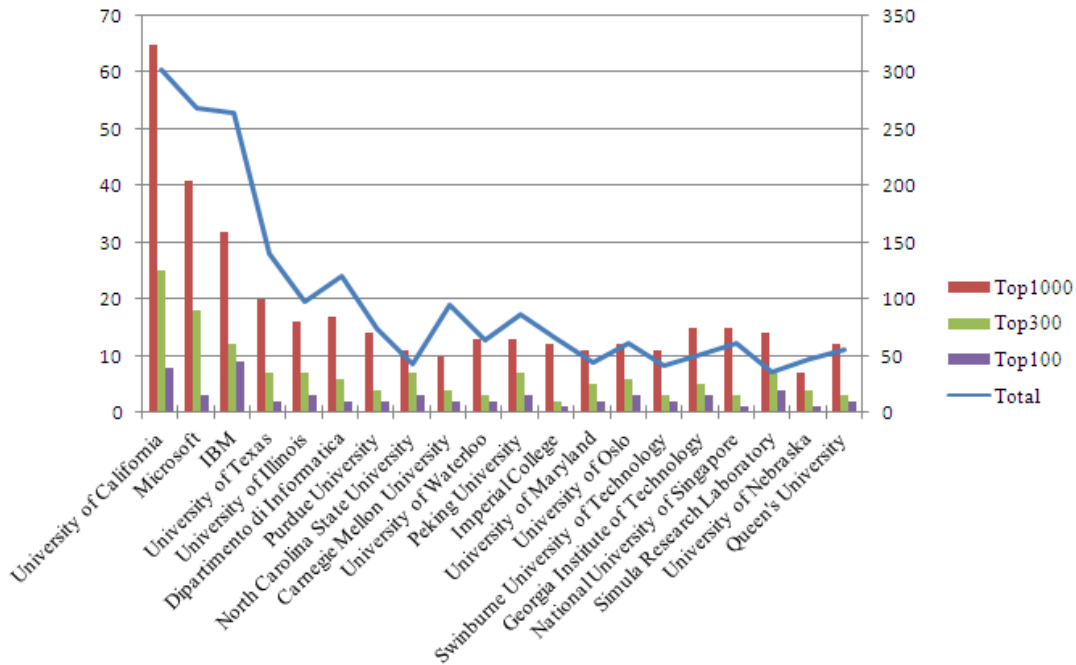

Figure 5. Composition of the leading institutions with different levels of scholars

An institution often has different levels of scholars, who are ranked as the Top 100, Top 300, or Top 1000 in terms of the $C$ rule. The composition of different levels of scholars in the Top 20 institutions is presented in Figure 5, where the vertical axes on the left and on the right represent the number of the Top-$k$ ($k$ = 100, 300 or 1000) scholars and the total number of scholars (who major in software engineering) in a leading institution, respectively.

As shown in Figure 5, the ranking of an institution is roughly proportional to the total number of affiliated scholars, that is, an institution seems to receive more scores if it has more scholars. However, there are two obvious exceptions, namely, North Carolina State University and Simula Research Lab. For the two institutions, the ratios of the number of the Top 100 and Top 300 scholars to the total number of scholars are higher than those of their neighbors, even though they have less affiliated scholars in software engineering. Interestingly, the proportion of the Top 1000 scholars to the total number of scholars is similar among all leading institutions. So, the number of the Top 100 or even Top 300 scholars has a relatively large impact on an institution's ranking, indicating that small-sized research institutions can also improve their rankings through raising affiliated scholars to be leading ones.

## 4.4 Leading Countries/Regions

The Top 20 published countries (or regions) are shown in Figure 6, where the numbers inside and outside the pie chart indicate a country/region' scores calculated in terms of the *C* rule and for its leadership respectively. According to the figure, these leading countries/regions have achieved a score that is not less than 643.5 during the years covered in this study, and most of the countries/regions' scores for leadership match their rankings, except China, Brazil, Norway and Hong Kong. The finding implies that these countries and regions have an obvious deficiency in leading the research in software engineering field, compared with their peers.

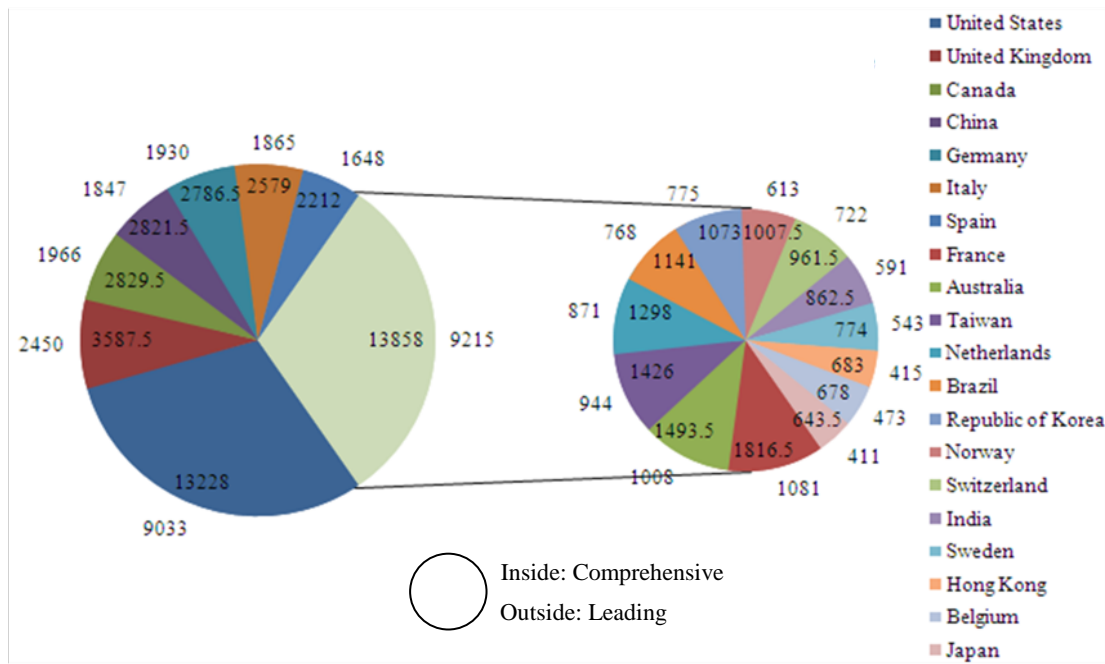

Figure 6. Top 20 countries or regions ranked in terms of the *C* rule

As shown in Figure 6, without doubt, the USA takes the first place of the list, and its score is more than three times the score of the runner-up, viz. the United Kingdom. For example, the Top 5 institutions (or even 8 out of the Top 10 institutions) are all in the USA (see Figure 4). Canada and China (not including Hong Kong, Taiwan and Macao) could be ranked for third place because of their similar scores, but there is a distinct gap between them and the United Kingdom. Among the leading countries/regions, most of them are from Europe (ten) and the Asia-Pacific region (seven). Besides, the USA and Canada are from North America, and Brazil is from South America.

## 4.5 Trend Analysis

As mentioned before, the trend analysis in this paper was conducted based on keywords, which were divided into two types. On one hand, the macro-keywords, derived from the EI classification terms, are used to reflect macro-level research subfields in software engineering and their changes. On the other hand, the micro-keywords, based on user-defined or uncontrolled keywords, are used to represent micro-level popular research topics and their changes.

### 4.5.1 Macro-research trends

Figure 7 and Figure 8 represent cumulated totals using scores and percentages each keyword contributes over time, respectively. These area charts drawn with the Microsoft Excel are used to show trends over time among related Top 20 macro-keywords, denoted by EI classification code and its corresponding classification term. Because the Top 20 macro-keywords are included, the first macro-keyword is plotted as a line with color fill followed by the second one, and so on. Note that, the Top 20 macro-keywords are shown in inversed order of their scores or percentages. That is, the one on the bottom takes first place, while the 20<sup>th</sup> macro-keyword is at the top of the list.

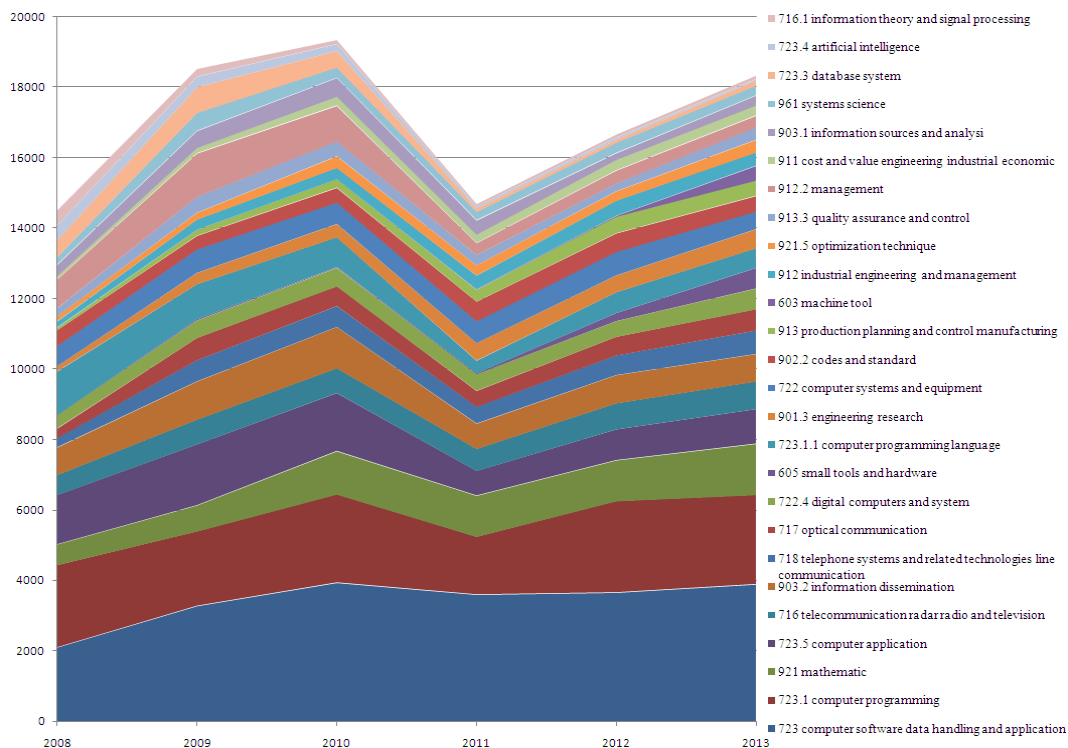

Figure 7. Trend analysis of macro-keywords (based on score)

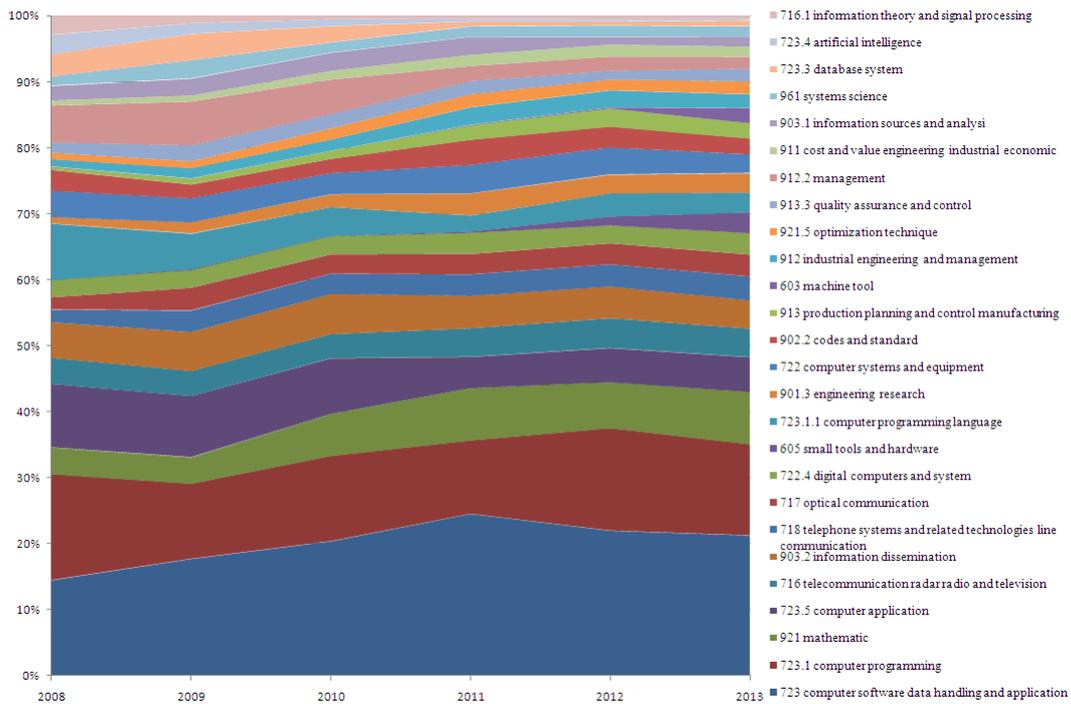

Figure 8. Trend analysis of macro-keywords (based on percentage)

According to the two figures, we find that (1) software engineering is an expanding research field driven by emerging application domains (e.g., telephone systems, telecommunication and optical communication) and the unceasing improvement of hardware and software technologies based on other mature disciplines such as mathematics, systems science and information theory; (2) a few macro-keywords that once were popular five years ago gradually become out of focus, e.g., management, database system, artificial intelligence, information theory and signal processing, while some macro-keywords are just the opposite over time, e.g., machine tool, production planning and control manufacturing, and small tools and hardware; (3) the core components of software engineering, such as computer programming, computer software data handling and application, computer application, and computer programming language, remain relatively stable in the past six years. The finding indicates that the principle sub-fields of software engineering changed little within the period analyzed, though new transitory application domains emerge and mature technologies of other disciplines go out little by little.

## 4.5.2 Micro-research trends

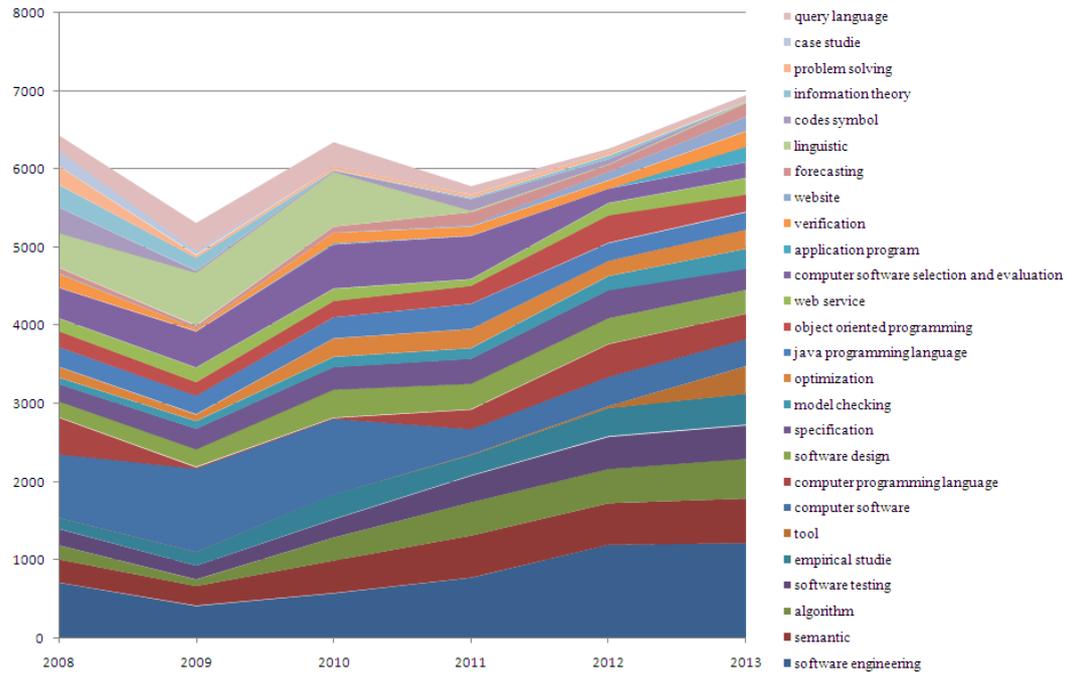

Figure 9. Trend analysis of micro-keywords (based on score)

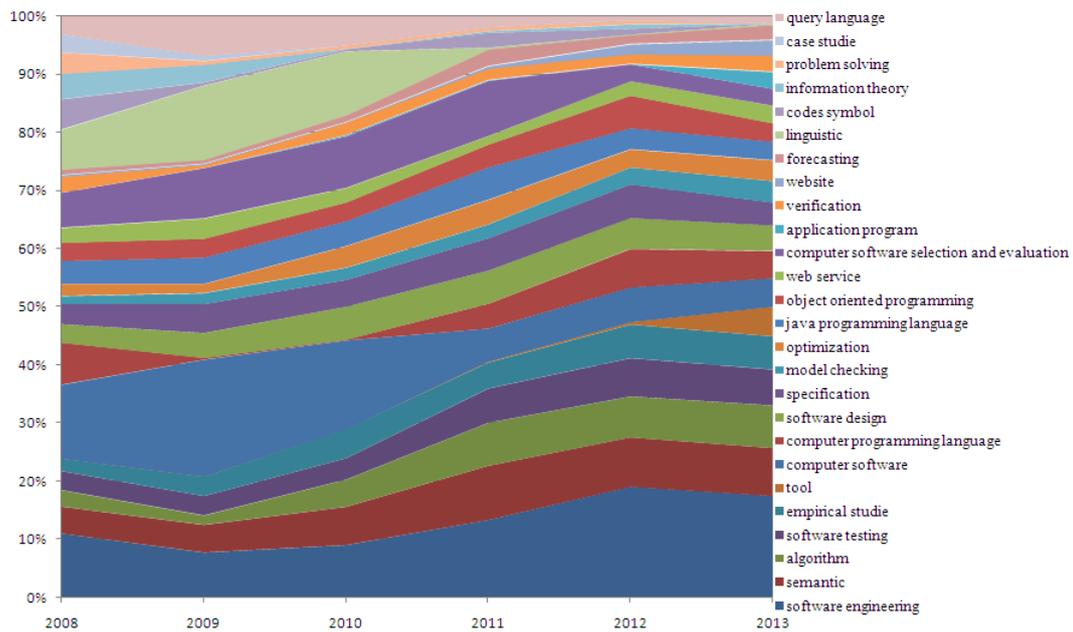

Figure 10. Trend analysis of micro-keywords (based on percentage)

The meanings of Figure 9 and Figure 10 are the same as Figure 7 and Figure 8 respectively. According to the two figures, we also find some interesting results described as follows.

(1) The term *Computer Software* took the first place of the list between 2008 and 2010, while the term *Software Engineering* climbed to the first in 2011. This implies that many researchers in the field of computer science still deemed software engineering or computer software as a research topic rather than a research field, though software engineering has grown into a burgeoning discipline compared with computer science. Therefore, researchers should tag their papers using more elaborated keywords instead of the general term *Software Engineering*.

(2) Since the micro-keywords are informally defined by authors, the changes of a few terms are more drastic than those of the macro-keywords (see Figure 10). Besides, most of them, which once were popular years ago, gradually descend to be unfashionable now, e.g., computer software, computer software selection and evaluation, linguistics and query language. Interestingly, the micro-keyword query language has the same trend change as its corresponding macro-keyword database system.

(3) Research topics, e.g., semantics, algorithm, software testing, empirical studies and software design, become popular and stable over time. This implies that software engineering attaches importance to both theoretical research (such as semantics and algorithm) and engineering practice (such as empirical studies). According to the ranking list in Figure 9 and Figure 10, software testing, software design and specification receive more attention than other phases in the whole software lifecycle. For software design, formal methods such as model checking and verification remain popular and stable in the past years; for software development, object-oriented programming and web services are still the mainstream development technologies.

## 4.6 Noteworthy Scholars and Institutions in Special Research Areas

Table 3. Noteworthy scholars and institutions in special research areas

| Research area | Noteworthy Scholars | Noteworthy Institutions |
|---|---|---|
| Computer software data handling and application | Harman, Mark; Poshyvanyk, Denys; Zhang, Lu; Kim, Sunghun; Mei, Hong | University of California; Microsoft; IBM |
| Computer programming | **Gulwani, Sumit**; Apel, Sven; Schäfer, Max; Visser, Eelco; Vitek, Jan | University of California; **Microsoft**; IBM |
| Computer application | Chan, W.K.; Tse, T.H.; Fraser, Gordon; Bodden, Eric; Chen, Tsong Yueh | University of California; Microsoft; IBM |
| Engineering research | **Jørgensen, Magne**; Kitchenham, Barbara; Budgen, David; Apel, Sven; Juristo, Natalia | University of Oslo; **Simula Research Laboratory**; Queen's University |
| Codes and standard | Maoz, shahar; Gabel, Mark; **Pradel, Michael**; Lo, David; Uchitel, Sebastian | **University of California**; **Imperial College London**; Microsoft |
| Production planning and control manufacturing | **Arcuri, Andrea**; Hemmati, Hadi; **Ali, Shaukat**; Dolby, Julian; Demsky, Brian | **University of California**; **IBM**; **Simula Research Laboratory** |
| Industrial engineering and management | **Gorschek, Tony**; Demsky, Brian; Krishnamurthy, Diwakar; Hemmati, Hadi; Arcuri, Andrea | IBM; **University of California**; **Blekinge Institute of Technology** |
| Optimization technique | Harman, Mark; Grunske, Lars; Masri, Wes; Tristan, Jean-baptiste; White, Jules | University of California; IBM; University of Texas |
| Quality assurance and control | Artzi, Shay; Grunske, Lars; Cataldo, Marcelo; De Lucia, Andrea; Yilmaz, Cemal | Microsoft; IBM; Queen's University |
| Cost and value engineering industrial economic | Mittas, nikolaos; **Angelis, Lefteris**; Do, Hyunsook; Hemmati, Hadi; **Briand, Lionel C.** | **University of Oslo**; Simula Research Laboratory; **Aristotle University** |

Table 3 presents the noteworthy scholars and institutions in ten special research areas within the field of software engineering, which were selected from the Top 20 macro-keywords according to the relevance to software engineering research and practice. If a scholar's name and the name of an institution are in bold type with the same color, the scholar is an affiliated member of the institution. Although the research areas are unable to cover scholars' main research interests, compared with the correlation between top institutions and top scholars in prior studies [20-22, 2], we argue that Table 3 (or an extension including more research areas) provides a useful "yellow pages" for young students or scholars in software engineering to search potential advisors or collaborators based on their referred papers published in a special research area, as well as to apply for those leading institutions based on its affiliated members' total contributions in the research area.

# 5 Discussion

## 5.1 Impact of Data Errors on the Results

Actually, the goal of the assessment and trend analysis of scientific research is to identify only a minority of the top ranking scholars, institutions and topics among all research subjects. For example, the prior studies in the field of software engineering only identify the Top15 scholars and institutions [7, 10-22, 2]. In this study, there are a total of 14,232 scholars and 4,638 institutions, but we focus mainly on the Top 20 scholars and institutions. According to the finding presented in the sub-section 4.1, the power-law distributions suggest that we don't need to consider the vast majority of research subjects in the long tail, because the difference between them and the leading ones is very distinct. Therefore, this is in good agreement with the goal of our study.

Data errors are unquestionably inevitable when we process data and conduct experiments. Because these errors arise randomly, we configure the software program with strict filtering rules in order to remove those mismatched research subjects. For example, if the name of an institution is written in a rare abbreviation that doesn't match its common ones, we neglect it without manual handling. So, the errors can result in a decrease in scores of a few research subjects. However, fixing such errors manually is time-consuming and costly. Then, we will present a theoretical proof that such errors in our study have negligible impact on the experimental results.

Assuming there are two instances $a$ and $b$ of a given type of research subject, their scores are $S_a$ and $S_b$ ($S_a > S_b$) respectively, and their rankings are $R_a$ and $R_b$ respectively. Data errors occur randomly in $n$ instances, leading to a decrease of total scores $S_{decrease}$. If $b$ climbs to (or even exceeds) the place of $a$, the probability that such errors cause a decrease of at least $S_a - S_b$ for $a$ is defined as

$$\frac{C_{S_a}^{S_a-S_b} \times C_{S_{all}-S_a}^{\alpha S_{all}-(S_a-S_b)}}{C_{S_{all}}^{\alpha S_{all}}}, \tag{1}$$

where $S_{all}$ is the total number of all instances' scores and $\alpha \in (0, 1)$ is a ratio of $S_{decrease}$ to $S_{all}$. It is worth noting that the real number scores in Formula 1 have been processed by the floor function.

For our discussion, if $R_a$ belongs to the Top 20 and $R_b$ is in the long tail, according to the properties of the power-law distributions of research subjects, $S_a - S_b$ is very closer to $S_a$, and Formula 1 can be simplified to

$$\frac{C_{S_{all}-S_a}^{\alpha S_{all}-S_a}}{C_{S_{all}}^{\alpha S_{all}}}. \tag{2}$$

Then, we used the definition of permutations and combinations [33] to expand the Formula 2. For simplify, we obtained Formula 3 described as follows.

$$\frac{(S_{all}-S_a)!}{(\alpha S_{all}-S_a)!\times(S_{all}-S_a-(\alpha S_{all}-S_a))!} \times \frac{(\alpha S_{all})!\times(S_{all}-\alpha S_{all})!}{S_{all}!}$$

$$= \frac{(S_{all}-S_a)!}{(\alpha S_{all}-S_a)!} \times \frac{(\alpha S_{all})!}{S_{all}!} = \frac{(S_{all}-S_a)!}{S_{all}!} \times \frac{(\alpha S_{all})!}{(\alpha S_{all}-S_a)!}$$

$$= \underbrace{\frac{\alpha S_{all}}{S_{all}} \times \frac{\alpha S_{all}-1}{S_{all}-1} \times \cdots \times \frac{\alpha S_{all}-S_a+1}{S_{all}-S_a+1}}_{S_a}$$

$$< \underbrace{\alpha \times \alpha \times \cdots \times \alpha}_{S_a} = \alpha^{S_a} \tag{3}$$

In fact, our study with careful data processing ensures that the value of $\alpha$ is less than 0.2. Besides, $S_a$ is larger than the mean score of all instances because of the top status of $R_a$. Hence, the value of the Formula 3 will be very small, indicating that the probability that a small amount of random errors has a great impact on the results is very small. That is to say, for example, our study can avoid the wrong conclusion that the Top 15[th] scholar and the first 300[th] scholar exchange their rankings, even though the occurrence of data errors is inevitable.

## 5.2  Selection of Data Source

Although the ACM Digital Library and the IEEE Xplore Digital Library are the two primary databases of full-text articles and bibliographic records in computer science, neither databases contain the data of all papers collected from the DBLP, implying that we have to collect experimental data from the two or more databases [31]. However, we found that this solution has the following disadvantages.

(1) These databases define different formats for author, institution, publisher, etc., and we have no choice but to develop a software program for heterogeneous data conversion with additional effort. Undoubtedly, this will cause more errors when processing experimental data.

(2) As we know, the ACM and IEEE databases have different classification systems. It is difficult to use the two different classification systems together to analyze macro-level research trends in software engineering, because sometimes they are contradictory.

In consideration of the above shortcomings of such a solution, we decided to select the EI Compendex database as the data source for our experiments, mainly because (1) it contains all referred papers published in the 36 publications, (2) it has unified data formats and a sole classification system, and (3) the data processing with the software tool *Octopus Collector* is highly cost-effective and time-efficient.

## 5.3  Disregarding Citation Number as an Evaluation Criterion

Some prior studies [13] considered that using the citation number of a paper as an evaluation criterion should be more reasonable for the assessment of a scholar's influence. Although we can easily get the citation number of each paper from the Scopus database[8], it is impractical and unreasonable for our study. The main reasons are described as follows.

(1) For some of the papers published between 2008 and 2013, especially between 2012 and 2013, the Scopus database has not yet collected their citation numbers when we tried to obtain them in January 2014.

(2) Some prior studies also argued that it is rational to evaluate the papers published at least five years ago in terms of citation number [34, 35], mainly due to the long period of peer review and publication of software engineering journals. In this study, the papers we examined were published in recent six years, so that it is not suitable to use citation number as an evaluation criterion.

## 5.4  Recognition of Differences among Publications

The journals examined in prior studies were deemed as equally-ranked peers regardless of their differences. However, most of scholars in software engineering may recognize that the IEEE TSE and the ACM TOSEM are more prestigious than the Springer ESE. On the other hand, the acceptance rates of few top conferences are very low, and their importance is not lower than most journals. The CCF considered the quality of 261 journals and 313 conferences in 10 important research areas in computer science, including systems and software engineering, and these publications were divided into three classes according to the existing lists of ranked journals and conferences as well as the collective discussion among native and foreign well-known scholars.

---
[8] http://www.scopus.com

To the best of our knowledge, neither the ACM nor the IEEE released such a ranking list. Based on the CCF's authority and influence, we believe that the ranking of publications is relatively objective and fair. Moreover, the evaluation rules in this paper have taken the quality of publications into consideration, so that our research results tend to be more reasonable than those of prior studies [7, 10-22, 2].

## 5.5 Threats to Validity

In this study, we obtained several interesting findings according to the proposed research questions. However, there still exist potential threats to the validity of our work.

Threats to *construct validity* are primarily related to the data we analyzed, which were collected from the EI Compendex database. We have explained the reason why the database was selected as our data source, though it is not a free database for open access. Therefore, we believe that our results based on the database are credible and can be reproduced.

Threats to *internal validity* are mainly related to the evaluation rules used in our study. For our experiments, we chose 36 different levels of publications and re-designed counting schemes for scholars, institutions and countries/regions based on those of the prior studies [**7**]. Hence, we are aware that the findings will certainly change if the number of publications, the rank of some publications, or the counting schemes vary. However, we have to admit that it is very hard to reach a widely recognized consensus on these factors for the assessment of scientific research.

Threats to *external validity* could be related to the generality of the method proposed in this paper. So, our future work is to apply it to other fields of computer science or other disciplines; besides, we will test the method with a questionnaire survey analysis of a large number of randomly selected scholars at home and abroad.

# 6   Conclusion

As we know, the assessment of scientific research is not a simple job. It is very hard to reach a widely recognized evaluation method for such a study. Although software engineering is a young discipline, the prior studies on the assessment of scholars and institutions have been reported. This paper presents a software-aided method for assessment and trend analysis, which can be used in software engineering as well as other research fields in computer science (or other disciplines).

The method proposed in this paper is modular and automated compared with the method in prior studies [7, 10-22, 2]. Besides, it takes into consideration more publications (including conference proceedings), the rank of each publication analyzed, and the different roles of authors in accomplishing a paper. According to the unified data source of the EI Compendex database, this paper presents two levels of research trend changes and those noteworthy scholars and institutions in a given research field, in addition to the assessment of scholars, institutions and countries/regions. Hence, we believe that the results could provide useful guidance on the selection of appropriate potential advisors or collaborators and the popular research topics in software engineering for newcomers or young scholars.

Our future work will focus primarily on applying this method to other research fields in computer science or other disciplines. On the other hand, we will improve the method with the feedback from randomly selected scholars involved in questionnaire surveys.

## Acknowledgement

This work is supported by the National Basic Research Program of China (No. 2014CB340401), the National Natural Science Foundation of China (Nos. 61273216 and 61272111), the Youth Chenguang Project of Science and Technology of Wuhan City in China (No. 2014070404010232), and the open foundation of Hubei Provincial Key Laboratory of Intelligent Information Processing and Real-time Industrial System (No. znss2013B017).